\documentstyle[aps,prb,twocolumn,epsf,floats]{revtex}

\begin{document}

\draft

\title{
  Reorientation transition of ultrathin ferromagnetic films
  }
\author{
  A.~Hucht\cite{email} and K.~D.~Usadel\cite{email2}
  }
\address{
  Theoretische Tieftemperaturphysik,
  Gerhard-Mercator-Universit\"at, 47048 Duisburg, Germany
  }
\author{
  \vskip -2\baselineskip\small(Received 20 August 1996)\break
  }
\author{
  \parbox{14cm}{\small
    \quad We demonstrate that the reorientation transition from out-of-plane
    to in-plane magnetization with decreasing temperature as observed
    experimentally in Ni-films on Cu(001) can be explained on a
    microscopic basis.  Using a combination of mean field theory and
    perturbation theory, we derive an analytic expression for the
    temperature dependent anisotropy.  The reduced magnetization in
    the film surface at finite temperatures plays a crucial role for
    this transition as with increasing temperature the influence of
    the uniaxial anisotropies is reduced at the surface and is
    enhanced inside the film.\hfill\break
    \hfill\break
    \leftline{PACS numbers:
      68.35.Rh, 75.10.Hk, 75.30.Gw, 75.70.-i
      }
    }
  }
\address{\vskip -1.5\baselineskip}


\maketitle


\markright{\rm
  {\it Physical Review} B {\bf 55}, 12309 (1997)
  }
\thispagestyle{myheadings}
\pagestyle{myheadings}

The direction of the magnetization of thin ferromagnetic films depends
on various anisotropic energy contributions like surface anisotropy
fields which often favor an orientation\cite{neel1} perpendicular to
the film,
dipole interaction which favors an in-plane magnetization,
and eventually anisotropy fields in the inner layers.
As a consequence of these competing effects, a temperature driven
reorientation transition from an out-of-plane ordered state at low
temperatures to an in-plane ordered state at high temperatures may be
observed at appropriate chosen film thicknesses.
Experimentally, this transition has been studied in
detail for various ultra-thin magnetic films\cite{allen1,pappas,qiu}.
Recently, it was found by Schulz and Baberschke\cite{bab1}
that ultra-thin Ni-films grown on Cu(001) show an opposite behavior:
the magnetization is oriented in-plane for low temperatures and
perpendicular at high temperatures.

Phenomenological approaches for explaining the reorientation
transition usually start from the energy (or the free energy 
at finite temperatures) which is expanded in terms of the orientation
of the magnetization vector relative to the film introducing
temperature dependent anisotropy coefficients $K_i(T)$.
The temperature dependence of these coefficients is then studied
experimentally (for a recent review see\cite{gradmann}). 

To better understand the mechanism responsible for the temperature
driven transition, several investigations have been done
in the framework of statistical spin models.
The advantage of this approach is that only a few microscopic
parameters enter: besides an exchange interaction the dipole
interaction and an uniaxial anisotropy in the surface layers of the
film. 
While Moschel {\it et al.}\cite{moschel1} showed that the temperature 
dependence of the reorientation transition is well described
qualitatively within a quantum mechanical mean field approach,
most other authors focused on classical spin models.
Extended Monte Carlo simulations on mono-layers\cite{chui,hucht1} as
well as mean field calculations of both mono-layers\cite{taylor}
and bilayers\cite{hucht2} agree in the sense that a temperature driven
reorientation transition is obtained.
Nevertheless, there is still a controversy with respect to the order
of this transition.
While Chui\cite{chui} measured the expectation value of the components
of the total magnetization and obtained a second order transition for
a monolayer
we found, using an improved simulation algorithm and analyzing the
Monte Carlo data with a histogram method, a transition of first order
in agreement with the mean field calculations for this
system\cite{hucht1}. 
Furthermore, we could show that the order of the transition depends on
the number of layers and on the distribution of the uniaxial
anisotropies\cite{hucht2}. 

In all of these theoretical investigations a temperature driven
reorientation transition from a out-of-plane state at low temperatures
to an in-plane state at high temperature is found for appropriate sets
of parameters, which is due to a competition of a positive surface
anisotropy and the dipole interaction.  The interesting new result for
ultra-thin Ni-films is argued \cite{bab1} to have its origin in a
stress-induced uniaxial anisotropy energy in the inner layers with its
easy axis perpendicular to the film. This anisotropy is in competition
with the dipole interaction and a negative surface anisotropy.  While
the thickness-dependent transition could be explained with these
anisotropies\cite{bab1}, the origin of the more interesting
temperature driven transition is not yet explained on a microscopic
basis. Note that the reversed reorientation recently found by MacIsaac
{\it et al.}\cite{macisaac} has a different origin as it only occurs
at vanishing exchange interaction.  It is the purpose of this letter
to show that the dipole interaction together with uniaxial
anisotropies in the film indeed may lead to a temperature driven
second order reorientation transition from an in-plane magnetized film
at low temperatures to a perpendicular magnetized film at high
temperatures.

After submission of this letter we became aware of a paper by Jensen
and Bennemann\cite{jb1} on the same topic. Starting from an expansion
of the free energy in terms of uniaxial anisotropy and dipole
interaction and employing then a mean field approximation following
earlier work\cite{levi} they calculated numerically the temperature of
both types of reorientation transitions. In contrast to this
calculation we develop in the present paper a fully selfconsistent
mean field theory and analyse the reorientation transition within this
approach since only within a nonlinear theory the canted phase and in
particular its width can be analyzed. Additionally, a selfconsistent
calculation of the quantities $K_i(\tau)$ introduced below which are
crucial for an understanding of the nature of the transition is not
possible. A linearization of the free energy which is discussed in the
last part of our paper agrees with the results of Ref.\cite{jb1}.
Within such a linearized theory the approximate location of the
transition can be obtained as a temperature somewhere within the
canted phase but it is not possible to calculate the width of this
phase. Thus, this calculation is only meaningful for a situation where
the canted phase occupies a rather small temperature interval. Note,
however, that this width can be quite large for instance for strongly
varying anisotropy energies or certain parameter
configurations\cite{hucht2} in which case a nonlinear approach is
necessary.

The calculations are done in the framework of a classical
ferromagnetic Heisenberg model consisting of $L$ two-dimensional
layers on a simple cubic lattice.
The Hamiltonian reads
\begin{eqnarray} \label{hami}
  {\cal H} & = & -{J \over 2}
  \sum_{\langle ij \rangle} {\vec s_i \cdot \vec s_j}
  - \sum_{i} {D_{\lambda_i} (s_i^z)^2} \nonumber\\
  & + & {\omega \over 2} \sum_{ij} { r_{ij}^{-3} \vec s_i \cdot \vec s_j
    - 3 r_{ij}^{-5} (\vec s_i \cdot \vec r_{ij})
    (\vec r_{ij} \cdot \vec s_j)},
\end{eqnarray}
where
\mbox{$\vec s_i = (s_i^x, s_i^y, s_i^z)$}
are spin vectors of unit length at position
\mbox{$\vec r_i = (r_i^x, r_i^y, r_i^z)$}
in layer $\lambda_i$ and
\mbox{$\vec r_{ij}=\vec r_i-\vec r_j$}.
$J$ is the nearest-neighbor exchange coupling constant,
$D_\lambda$ is the uniaxial anisotropy which depends on the layer
index \mbox{$\lambda = 1 \ldots L$}, and
\mbox{$\omega = {\mu_0 \mu^2}/{4 \pi a^3}$}
is the strength of the long range dipole interaction on a lattice with
lattice constant $a$ ($\mu_0$ is the magnetic permeability and $\mu$
is the effective magnetic moment of one spin). 
All energies and temperatures are measured in units of $J$
($k_{\text B} = 1$) which is fixed to \mbox{$J = 1$} in this letter.
Note that only second order uniaxial anisotropies $D_\lambda$ enter
the Hamiltonian Eq.~(\ref{hami}).
In our calculations we will restrict ourself to the case that all
anisotropies are the same except at one surface, as this scenario is
sufficient for explaining the basic physics of the temperature driven
reorientation transition. Furthermore we will focus on the case
of $L=4$ layers.
A systematic investigation of the parameter and thickness dependence of the
reorientation transition and in particular a calculation of the 
corresponding phase diagrams is under way\cite{hucht4}.

\begin{figure}[t]
  \epsfxsize=8cm
  \epsffile{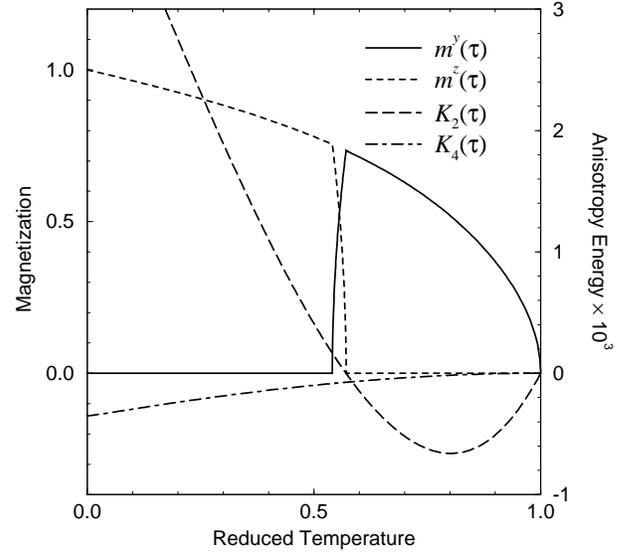}
  \caption{
    Magnetization components and anisotropy energies for a Fe-type
    system with $L = 4$ layers.
    $J = 1$,
    $D_1 / J       = 14 \times 10^{-3}$,
    $D_{\lambda>1} = 0$,
    $\omega / J    = 38 \times 10^{-5}$.
    The parameters are based on Iron.
    \label{f:fe}
    }
\end{figure}
In the following we assume translational invariance within the layers
and therefore we set
\mbox{$\langle\vec s_i\rangle = \vec m_\lambda$}
if $\vec s_i$ is a spin in layer $\lambda$.
For the Hamiltonian Eq.~(\ref{hami}) a molecular-field approximation is
implemented resulting in $L$ effective one particle Hamiltonians from
which the free energy functional can be obtained:
the mean field in layer $\lambda$ is given by
\mbox{$\vec h_\lambda = \sum_\mu{{\bf X}_{\lambda \mu} \vec m_\mu}$}
where ${\bf X}_{\lambda \mu}$ contains both exchange and dipole
interaction. With the order parameter
\mbox{${\bf M} = (\vec m_1,...,\vec m_L)$}
the Hamiltonian in layer $\lambda$ becomes 
\begin{equation} \label{hmf}
  {\cal H}_\lambda^{MF}({\bf M})
  = \vec h_\lambda \cdot
  \left( {1 \over 2} \vec m_\lambda - \vec s_\lambda \right)
  - D_\lambda (s_\lambda^z)^2.
\end{equation}
Integrating this mean field Hamiltonian over the surface of the unit
sphere in each layer yields the free energy per surface
element, 
\begin{equation} \label{frei1}
  {\cal F}(T, {\bf M})
  = - T \sum_{\lambda=1}^{L}{\log\oint{
      d\vec s_\lambda e^{- {\cal H}_\lambda^{MF}({\bf M}) / T}}}.
\end{equation}
Due to the in-plane rotational invariance of Eq.~(\ref{hmf})
we can set \mbox{$m_\lambda^x = 0$} and thus the free energy in
Eq.~(\ref{frei1}) depends on the $2 L$ components $m_\lambda^y$ and
$m_\lambda^z$ of ${\bf M}$ and is stationary with respect to
variations of these quantities. 
This variation is done in two steps:
First we minimize the free energy Eq.~(\ref{frei1}) with the constraint
that the azimuth angle $\vartheta$ of the total magnetization
\mbox{$\vec m = L^{-1} \sum_\lambda{\vec m_\lambda}$}
is fixed, and expand the resulting constrained free energy in powers
of $\cos(\vartheta)$ to give the angle-dependent free energy
\begin{eqnarray} \label{frei2}
  {\cal F}(\tau, \vartheta) & = & {\cal F}_0(\tau)
  -    K_2(\tau) \cos^2(\vartheta) \nonumber\\
  && - K_4(\tau) \cos^4(\vartheta) - ...
\end{eqnarray}
with the reduced temperature
\mbox{$\tau = T/T_c$} ($T_c$ is the Curie temperature of the film)
and temperature dependent expansion coefficients $K_i(\tau)$.
These quantities are usually introduced phenomenologically. However in
our approach we can calculate these coefficients $K_i(\tau)$ from the
microscopic parameters of the system.
The equilibrium free energy is then obtained as the minimum of
Eq.~(\ref{frei2}) with respect to $\vartheta$.

In this notation, the two reorientation transition temperatures
$\tau_r^z$, where \mbox{$m^z \rightarrow 0$},
and
$\tau_r^y$, where \mbox{$m^y \rightarrow 0$},
are given by the conditions
\begin{mathletters}
  \begin{eqnarray}
    0 & = & K_2(\tau_r^z), \\
    0 & = & K_2(\tau_r^y) + 2 K_4(\tau_r^y).
  \end{eqnarray}
\end{mathletters}
Fig.~\ref{f:fe} shows the temperature dependence of the components of
the total magnetization $\vec m(\tau)$ and the anisotropy coefficients
$K_2(\tau)$ and $K_4(\tau)$ for a situation where one of the layers
(the surface layer) has a positive uniaxial anisotropy,
\mbox{$D_1 > 0$}, and the others are set to
\mbox{$D_{\lambda>1} = 0$}.
$K_6(\tau)$ and higher order terms are nearly two magnitudes smaller than
$K_2(\tau)$ and therefore are not depicted.
This is the situation encountered for instance in ultra-thin Fe-films.
The ground-state magnetization of the system is perpendicular to the
film.
Increasing the temperature the magnetization switches
continuously from the perpendicular direction at a temperature
$\tau_r^y$ to the in-plane direction at $\tau_r^z$.

For Ni-films on Cu(001) there is a positive uniaxial volume anisotropy
in the inner layers favoring perpendicular orientation, and eventually
a negative anisotropy
on the surface\cite{bab1}. In competition to these energies is the
dipole interaction which always favors in-plane magnetization.
\begin{figure}[t]
  \epsfxsize=8cm
  \epsffile{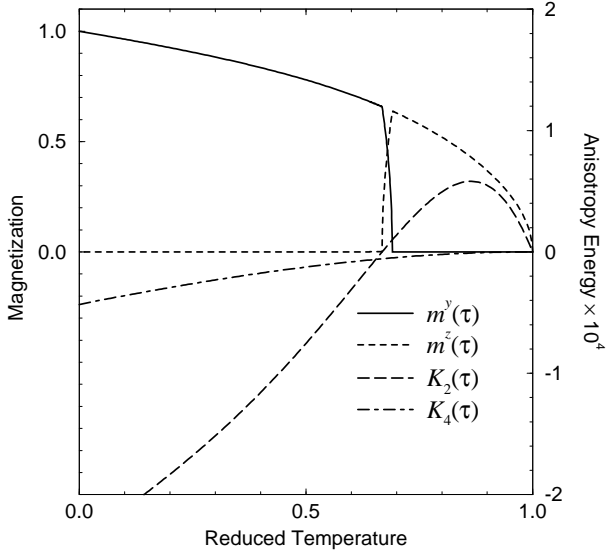}
  \caption{
    Magnetization components and anisotropy energies for a Ni-type
    system with $L = 4$ layers.
    $J = 1$,
    $D_1 / J           = -3.5 \times 10^{-3}$,
    $D_{\lambda>1} / J =  1.5 \times 10^{-3}$,
    $\omega / J        =  5   \times 10^{-5}$.
    The parameters are based on Nickel.
    \label{f:ni}
    }
\end{figure}
Fig.~\ref{f:ni} shows the temperature dependence of the components
of the total magnetization vector and the anisotropy coefficients
$K_i(\tau)$ for a Ni-type system with \mbox{$L = 4$} layers. The
exchange interaction $J$ is estimated from the Curie temperature of
bulk Nickel, the dipole constant $\omega$ is calculated from the ground
state magnetic moment and the lattice constant, and the anisotropy
energies are taken from the experiment\cite{bab1}.
For these parameters with increasing temperature the magnetization
starts to cant at a temperature $\tau_r^z$ and reaches the
perpendicular state at $\tau_r^y$ as observed experimentally in
Ni-films. These results were obtained numerically by solving the
corresponding mean field equations.

In order to understand both the normal and the reversed reorientation
transition we additionally applied a perturbation theory to the mean field
Hamiltonian Eq.~(\ref{hmf}) considering $\omega$ and $D_\lambda$ as small
perturbations of the pure isotropic Heisenberg Hamiltonian which is
justified in view of the smallness of these parameters.
We will only give the results of these calculations in this letter,
the complete derivation will be reported in detail in a forthcoming
paper\cite{hucht4}. 

The total anisotropy $K(\tau)$ of the system is defined as the
difference of the free energies of the in-plane state and the
out-of-plane state
\begin{eqnarray} \label{ktot1}
  K(\tau) & = & {\cal F}(\tau,\pi/2) - {\cal F}(\tau,0)
  \nonumber\\
  & = & K_2(\tau) + K_4(\tau) + ...~.
\end{eqnarray}
When we neglect the narrow canted phase, the reorientation
temperature $\tau_r$ is given by the condition
\mbox{$K(\tau_r) = 0$}.
If the first derivative
\mbox{$\partial_\tau K(\tau_r) < 0$},
we have a normal transition from out-of-plane to in-plane
magnetization direction, otherwise the transition is reversed.

In the framework of a perturbation theory we can derive an analytical
expression for $K(\tau)$ involving the absolute value of the layer
magnetizations $m_\lambda(\tau)$ and the fluctuations transversal to the
magnetization direction
\mbox{$q_\lambda(\tau)=\langle(s_\lambda^\perp)^2\rangle(\tau)$},
both calculated with the unperturbed Hamiltonian:
\begin{eqnarray} \label{ktot2}
  K(\tau) & = & K_q(\tau) + K_m(\tau)
  \nonumber\\
  & = & \sum_{\lambda=1}^L{
    D_\lambda \left( 1 - 3 q_\lambda(\tau) \right)}
  \nonumber\\
  & & - {3 \omega \over 4} \sum_{\lambda,\lambda'=1}^L
  {m_\lambda(\tau) \Phi_{|\lambda-\lambda'|} m_{\lambda'}(\tau)}.
\end{eqnarray}
The constants $\Phi_\delta$ contain the effective dipole
interaction between the layers and can be calculated numerically to give
\mbox{$\Phi_0 = 9.0336$}, \mbox{$\Phi_1 = -0.3275$}\cite{error}, and 
\mbox{$\Phi_{\delta>1} = {\cal O}(e^{-2 \pi (\delta-1)})$}.

At the critical temperature $K(\tau)$ vanishes and hence $K(\tau)$
must be curved in order to have another zero at a temperature
$\tau_r<1$.
Furthermore, a positive curvature is necessary for a normal
reorientation transition while a negative curvature of $K(\tau)$ is
necessary for a reversed reorientation.
Hence we will focus on the second derivatives of
Eq.~(\ref{ktot2}) and start with the dipole part $K_m(\tau)$.
It turns out that \mbox{$\partial_\tau^2 K_m(\tau) > 0$}
for all film thicknesses and temperatures since $\omega$ is positive and
the main contribution of the sum is proportional to
$\sum_\lambda{m_\lambda^2(\tau)}$ which always has a negative curvature.
Thus the dipole interaction always favors the normal reorientation and
can never lead to a reversed transition in an exchange dominated
system. 

Now we will examine $K_q(\tau)$.
First note that $q_\lambda(0)=0$ and $q_\lambda(1)=1/3$ in the
unperturbated case.
For $L=1$ and $L=2$ we have \mbox{$q_\lambda(\tau) = \tau/3$}
in mean field approximation, and then
\mbox{$K_q(\tau) = \sum_\lambda{D_\lambda} (1-\tau)$},
i.e.~the second derivative vanishes in this case.
Consequently in systems with $L\le2$ layers we only find normal
reorientation transitions from out-of-plane direction at low
temperatures to an in-plane direction at higher temperatures.

This is not the case for $L>2$ layers since then, due to the reduced
surface magnetization at finite temperatures, the transversal
fluctuations in the surface layers $q_s(\tau)$ are enhanced
($q_s(\tau) \ge \tau/3$,
\mbox{$\partial_\tau^2 q_s(\tau) < 0$}).
Combined with a negative surface anisotropy $D_s$
this may lead to a negative curvature of $K_q(\tau)$.
Furthermore, the transversal fluctuations in the inner layers are
reduced by this effect 
($q_b(\tau) \le \tau/3$,
\mbox{$\partial_\tau^2 q_b(\tau) > 0$}) and, when combined with a
positive uniaxial anisotropy in the inner layers enhance the
negative curvature of $K_q(\tau)$.

\begin{figure}[t]
  \epsfxsize=8cm
  \epsffile{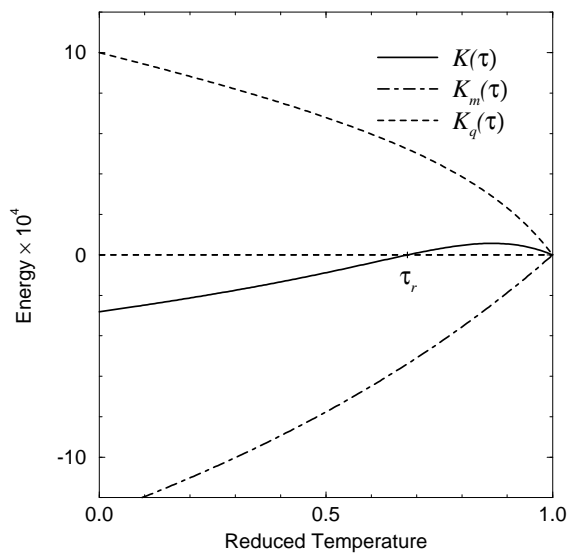}
  \caption{
    Total anisotropy $K(\tau)$ and its two parts
    $K_q(\tau)$ and $K_m(\tau)$ from Eq.~\ref{ktot2}
    for a Ni-type system.
    The model parameters are the same as in Fig.~\ref{f:ni}.
    \label{f:df}
    }
\end{figure}
In Fig.~\ref{f:df} $K(\tau)$ is depicted together with the two competing
parts $K_q(\tau)$ and $K_m(\tau)$ from Eq.~(\ref{ktot2}) for the same
parameters as in Fig.~\ref{f:ni}. A transition is obtained with
increasing temperature because $K_q(\tau)$ tends slower to zero than
$K_m(\tau)$.

In summary we have shown that the temperature driven reorientation transitions
seen in ultra-thin ferromagnetic films are well described within a
mean field approximation if second order uniaxial anisotropies and the
dipole interaction are included in the Hamiltonian.
In particular we can relate the unusual transition seen in Ni-films to
a microscopic model in which a positive uniaxial anisotropy energy is
present in the inner layers. 
Additionally we can calculate the parameters $K_i(T)$ usually
introduced phenomenologically from microscopic parameters of the
system.

The $L=4$ layer film considered serves as a simple system showing the
Ni-type transition, while the Fe-type transition is already observed
in mono layers if the parameters are adjusted properly.
This has a rather interesting physical origin:
For systems with $L=1$ or $L=2$ layers the unperturbed system is
homogeneous as
every lattice site has the same environment.
It turns out that in this case only a reorientation transition from
out-of-plane to in-plane can occur, provided the exchange interaction
is large with respect to the uniaxial anisotropies and the dipole
interaction.
When the film thickness $L > 2$, the magnetization is not homogeneous
through the film as the surface layers have a reduced magnetization
at finite temperatures.
This leads to an enhancement of the transversal fluctuations at the
surface and to a reduction of these fluctuations in the inner of the
film.
Hence the influence of the uniaxial anisotropies is reduced at the
surface and enhanced inside the film favoring a spin orientation
parallel to the easy axis of the
inner layer.
This effect may lead to a temperature driven reorientation
transition of the type observed in Nickel.

\vspace{6pt}
This work was supported by the Deutsche Forschungsgemeinschaft through
Sonderforschungsbereich 166. The authors would like to
thank S.~L\"ubeck and A.~Moschel for fruitful discussions.



\begin{references}
\bibitem[\star]{error} Erratum: In the publication this constant has
  the wrong value \mbox{$\Phi_1 = -0.65493$}.
\end{references}
\end{document}